\documentclass[12pt,psfig,epsfig,twoside]{article}
\usepackage{fleqn,psfig,epsf,espcrc1}
\usepackage{graphicx}

\newcommand{\AmS}{{\protect\the\textfont2
  A\kern-.1667em\lower.5ex\hbox{M}\kern-.125emS}}

\def\shiftdown#1{#1\llap{\lower.04ex\hbox{#1}}}

\begin{document}

\vspace*{-48pt}
\begin{center}
{\large\bf
Relativistic field-theoretical approach to the
inverse scattering problem.
}
\end{center}
\vspace{1cm}
\noindent{\em A.I.\ Machavariani}

\vspace{3pt}
\noindent{ Joint\ Institute\ for\ Nuclear\ Research,\ Dubna,\ Moscow\
region\ 141980,\ Russia\\
and High Energy Physics Institute of Tbilisi State University,
University str.  9 }\\
{\em Tbilisi 380086, Georgia }

\vspace{1.5cm}

The inverse scattering problem for the relativistic 
three-dimensional equation\\
$\Bigl(2E_{\bf p'}-2E_{\bf p}\Bigr)<{\bf p'}|\Psi_{\bf p}>=
-\int V(t)d^3{\bf p''}<{\bf p''}|\Psi_{\bf p}>$ with
$E_{\bf p}=\sqrt{m^2+{\bf p}^2}$ and $t=\Bigl(E_{\bf p'}-E_{\bf p''}\Bigr)^2
-\Bigl({\bf p'}-{\bf p''}\Bigr)^2$ is considered.
The field-theoretical potential $V(t)$ of this equation is constructed
in the framework of the old perturbation theory. It contains all 
contributions of diagrams with intermediate off-mass shell particles.
In particular,  this potential reproduces the OBE
 Bonn model of the $NN$ potential exactly.
For the $\pi N$ scattering it 
is generated by $\sigma,\rho, \omega$-meson exchange diagrams.
The inverse scattering problem is solved by reduction of these
relativistic equations to the standard Schr\"odinger equations 
$\Bigl(\Delta_{\bf r}+{\bf k}^2\Bigr)<{\bf r}|\phi_{\bf p}>=
- v({\bf r})<{\bf r}|\phi_{\bf k}>$ with
$E_{\bf p}={\bf k}^2/2m+m$. The relation between the relativistic
potential
$V(t)$ and its nonrelativistic representation $v({\bf r})$ is obtained.

\newpage

The inversion method for the reconstruction of  the potential in the
 Schr\"odinger equation is needed for the numerous
 problems in nuclear and particle physics \cite{Chad,New,Suz,Ger}.
However, the relativistic generalizations of the inverse scattering methods
in the quantum scattering theory was carried out  for the one particle
relativistic equations such as the Dirac or Klein-Gordon equation only.
\cite{Chad}.
The aim of  this paper is to extend the inversion method
for the  determination of  
the potential in the
field-theoretical equations. In particular we consider
the scheme of the relativistic generelization of the Schr\"odinger
equations for the $NN$ scattering amplitudes\cite{Ger2}.

 The relativistic
equation 
in the time-ordered three-dimensional
field-theoretical formulation for the
$NN$ amplitude $<{\bf p}'|t\Bigl(E({\bf p})\Bigr)|{\bf p}>$ \cite{MCH,M1}
 has the form of the Lippmann-Schwinger equation

$$
<{\bf p}'|t\Bigl(E({\bf p})\Bigr)|{\bf p}>=
<{\bf p}'|U\Bigl(E({\bf p})\Bigr)|{\bf p}>
+\int d{\bf p}''
{{<{\bf p}'|U\Bigl(E({\bf p})\Bigr)|{\bf p}''>}
\over{E({\bf p})+io-E({\bf p}'')}}
<{\bf p}''|t\Bigl(E({\bf p})\Bigr)|{\bf p}>, \eqno(1)$$
where ${\bf p}'$ and ${\bf p}$ denotes the nucleon relative momentum in the
c.m. frame and $E({\bf p})=2E_{\bf p}\equiv 2\sqrt{{\bf p}^2+m_N^2}$. 

The relativistic equation (1) for the $NN$ amplitude was derived in
the framework of    
the standard $S$-matrix reduction technique by  using the completeness condition
 for the asymptotic $''in''$ states \cite{MCH,M1}. Moreover in
Ref.\cite{M1} there were considered the explicit connections of the equation (1) 
with the quasipotential reduction of the Bethe-Salpeter equation 
and with the other three-dimensional
time-ordered field-theoretical equations. The brief analysis of the
Lippmann-Schwinger type equation (1) is given in the appendix.

The potential in the equation (1) consist of the on-mass shell
meson exchange potential (A6c) (Fig.1) and   of the $NN$ potential
${\cal Y}(t)$ (A6a,b) which are generated by the equal-time commutators
and which  dependent on the square of the four-momentum transfer $t$.
The $NN$ potential ${\cal Y}(t)$ contains  the 
off-mass shell meson exchange part (Fig.2A) and  the $NN$
overlapping (contact) potential  (Fig.2B).   

\vspace{3mm}

\begin{figure}[htb]
\centerline{\epsfysize=145mm\epsfbox{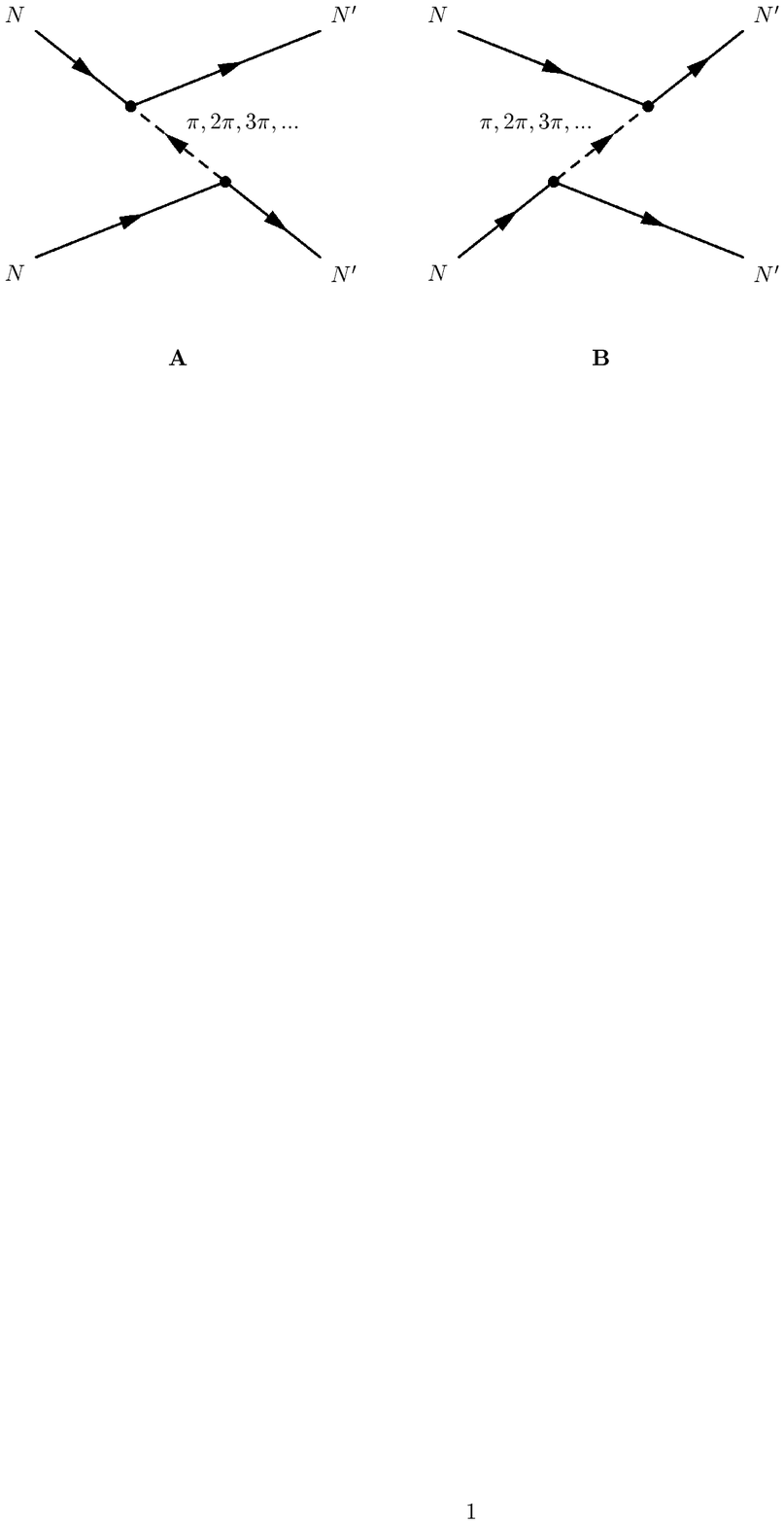}}
\vspace{-11.0cm}
\caption {\footnotesize {\it Diagrammatic representation of 
the time-ordered $NN$ 
interaction potential  with on-mass shell intermediate pions. 
The  full circles denote the vertex functions  
$
<{\bf p'}_N|J_{{\bf p}_N}(0)|{\bf p}_{\pi 1},{\bf p}_{\pi 2}...;in>$ and 
$<0|{\overline J}_{{\bf p}_N}(0)|{\bf p}_N{\bf p}_{\pi 1}
{\bf p}_{\pi 2}...;in>$
with the one off-mass shell nucleon.}}
\label{fig:one}
\end{figure}

\vspace{3mm}

\begin{figure}[htb]
\centerline{\epsfysize=145mm\epsfbox{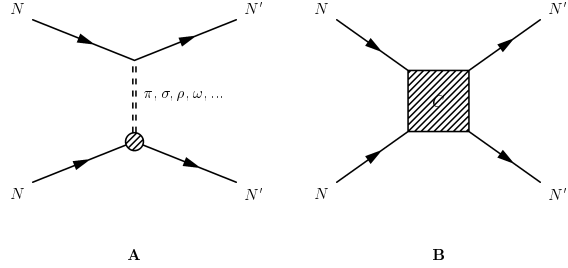}}
\vspace{-11.0cm}
\caption {\footnotesize {\it 
The $NN$ interaction with the off-mass shell $\pi,\sigma,
\rho,\omega,...$ meson exchange-diagram (${\bf A}$)
and with the four nucleon overlapping or contact term diagram  
 (${\bf B}$). The shaded circle corresponds to the vertex function 
$<{\bf p'}_N|j_{\pi}(0)|{\bf p}_N>$ with off-mass shell pi-meson and
the other $\pi NN$
vertex functions in the diagram ${\bf A}$
are given in the tree approximation according to equation (4).
}}
\label{fig:two}
\end{figure}
\vspace{3mm}

The vertex functions   
$<{\bf p'}_N|J_{{\bf p}_N}(0)|{\bf p}_{\pi 1}{\bf p};in>$ and 
$<0|{\overline J}_{{\bf p}_N}(0)|{\bf p}_N{\bf p}_{\pi 1};in>$ of the 
transition $N\to N\pi_1$
can be determined from the $\pi N$ scattering phase shifts by
using the dispersion relations \cite{BBB}.
The vertex functions  
$<{\bf p'}_N|J_{{\bf p}_N}(0)|{\bf p}_{\pi 1}{\bf p}_{\pi 2};in>$ and 
$<0|{\overline J}_{{\bf p}_N}(0)|{\bf p}_N{\bf p}_{\pi 1}{\bf p}_{\pi 2};in>$
of the transition $N\to N\pi_1\pi_2$
can be obtained by using the results of the calculation of 
the $\pi N$ scattering amplitude
with one crossed pion etc.
Therefore we can assume that the nonlocal, on-mass shell meson exchange
potential in eq.(A6c)-see (Fig.1) is defined ``a priory''. Afterwards  
one can subtract   the contributions of
this ``a priory''  fixed potential from the complete $NN$ phase shifts.
Thus the inverse scattering problem for the relativistic
Lippmann-Schwinger type equation (1) is reduced to the investigation 
of the  relativistic equation 

$$
<{\bf p}'|{\cal T}|{\bf p}>=
{\cal Y}\biggl( (E_{{\bf p}'}-E_{{\bf p}})^2-({\bf p'-p})^2\biggr)$$
$$+\int 
{\cal Y}\biggl( (E_{{\bf p}'}-E_{{\bf p''}})^2-({\bf p'-p''})^2\biggr)
{{d{\bf p}''}\over{2E_{\bf p}+io-2E_{{\bf p}''} }}
<{\bf p}''|{\cal T}|{\bf p}>,\eqno(2)$$
where $E_{\bf p}=\sqrt{m_N^2+{\bf p}^2}$ and the $NN$ potential
${\cal Y}$ (A6b,c) is depicted on the Fig.2.

It is convenient to present equation (2) for the wave function 
$<{\bf p}'|\Psi_{\bf p}>$

$$\Bigl(2E_{\bf p'}-2E_{\bf p})<{\bf p}'|\Psi_{\bf p}>=
\int d{\bf p}''
{\cal Y}\biggl( (E_{{\bf p}'}-E_{{\bf p''}})^2-({\bf p'-p''})^2\biggr)
<{\bf p}''|\Psi_{\bf p}>,\eqno(3)$$
where 
$<{\bf p}'|{\cal T}|{\bf p}>=<{\bf p}'|{\cal Y}|\Psi_{\bf p}>$.

In  order to transform  Eq.(3) to the  
nonrelativistic form we introduce the variables

$$ k^2=m_N(2E_{\bf p}-2m_N);\ \ \  {{\bf k}\over k}={ {\bf p}\over p}
\eqno(4)$$
and
$${\bf u}={\bf k}'\sqrt{{{k''^2}\over{4m_N^2}}+1};
 \ \ \ \ \ 
{\bf v}={ {\bf k}''}\sqrt{{{{k'}^2}\over{4m_N^2}}+1}
\eqno(5)$$
which satisfy the important condition {\footnotemark}

\footnotetext{ For the variables ${\bf u}$ and ${\bf v}$ 
there is a  more transparent representation 
${\bf u}=
{\bf p}'\sqrt{ {E_{{\bf p}'}+m_N}\over {E_{{\bf p}''}+m_N} }$ and
${\bf v}={\bf p}''
\sqrt{ {E_{{\bf p}''}+m_N}\over {E_{{\bf p}'}+m_N} }$.
}

$$t=-({\bf u}-{\bf v})^2=(E_{{\bf p}'}-E_{{\bf p}''})^2-({\bf p'-p''})^2.
\eqno(6)$$

Then equation (3) takes the form

$$({\bf k'}^2-{\bf k}^2)<{\bf k}'|\Psi_{\bf k}>=
\int d{\bf k}''
 Y\biggl(-({\bf u}-{\bf v})^2\biggr)
<{\bf k}''|\Psi_{\bf k}>,\eqno(7)$$
where 
$J^{1/2}(k'')<{\bf k}''|\Psi_{\bf k}>=<{\bf p}''|\Psi_{\bf p}>$;
$J^{1/2}(k') Y\biggl(-({\bf u}-{\bf v})^2\biggr)J^{1/2}(k'')={\cal Y}(t)$
and $J(k'')={p''}^2dp''/{k''}^2dk''$.

Now we present Eq. (7) in the coordinate space using
the Fourier transform

$$\Bigl(\Delta_{\bf r'}+{\bf k}^2)<{\bf r}'|\Psi_{\bf k}>=
-\int e^{-i{\bf k'r'}}{{d{\bf k}'}\over{(2\pi)^3}}{{d{\bf k}''}\over{(2\pi)^3}}
 Y\biggl(-({\bf u}-{\bf v})^2\biggr)
e^{i{\bf k''r''}}d{\bf r''}
<{\bf r}''|\Psi_{\bf k}>\eqno(8)$$

Inserting

$$  Y\biggl(-({\bf u-v})^2\biggr)=\int d{\bf z}e^{i({\bf
u-v}){\bf z} }  Y({\bf z}).\eqno(9)$$ 

into Eq. (8) we obtain

$$\Bigl(\Delta_{\bf r'}+{\bf k}^2)<{\bf r}'|\Psi_{\bf k}>=
-\int \biggl\{
e^{{\bf z}\nabla_{\bf r'}[\sqrt{\Delta_{\bf r''}/4m_N^2+1}-1]}
e^{-{\bf z}\nabla_{\bf r''}[\sqrt{\Delta_{\bf r'}/4m_N^2+1}-1]}$$
$$
e^{-i{\bf k'r'}}e^{i{\bf k''r''}}
e^{i{\bf k'z}}e^{-i{\bf k''z}}\biggr\}
d{\bf r''}
{{d{\bf k}'}\over{(2\pi)^3}}{{d{\bf k}''}\over{(2\pi)^3}}
  Y({\bf z})d{\bf z}
<{\bf r}''|\Psi_{\bf k}>, \eqno(10)$$
where the operators in the big curly brackets act on the function
which are included in this brackets and  
$\Delta_{\bf r'}=\nabla_{\bf r'}^2\equiv \partial_{\bf r'}^2$  

After integration over ${\bf k}'$ and ${\bf k}''$ we find 

$$\Bigl(\Delta_{\bf r'}+{\bf k}^2)<{\bf r}'|\Psi_{\bf k}>=
-\biggl\{
\int e^{{\bf z}\nabla_{\bf r'}[\sqrt{\Delta_{\bf r''}/4m_N^2+1}-1]}
e^{-{\bf z}\nabla_{\bf r''}[\sqrt{\Delta_{\bf r'}/4m_N^2+1}-1]}
$$
$$d{\bf r''}
\delta({\bf z-r'})\delta({\bf r''-z})\biggr\}
  Y({\bf z})d{\bf z}
<{\bf r}''|\Psi_{\bf k}>.\eqno(11)$$

The  operators
$e^{{\bf z}\nabla_{\bf r'}[\sqrt{\Delta_{\bf r''}/4m_N^2+1}-1]}$
and $e^{-{\bf z}\nabla_{\bf r''}[\sqrt{\Delta_{\bf r'}/4m_N^2+1}-1]}$
compensate each other on the surface
${\bf r'=r''=z}$. Therefore we obtain the standard Schr\"odinger equation

$$
\Bigl(\Delta_{\bf r'}+{\bf k}^2)<{\bf r}'|\Psi_{\bf k}>=
 - Y({\bf r}')<{\bf r}'|\Psi_{\bf k}>\eqno(12) $$

Our main result is
the explicit reduction of the relativistic Lippmann-Schwinger equation (3)
with the potential ${\cal Y}(t)$ to the standard Schr\"odinger
equation (12) with the auxiliary variable
$ 2m_N\sqrt{{\bf p}^2+m_N^2}-2m_N^2={\bf k}^2$ (4). 
This result allows to determine 
 the relativistic potential
${\cal Y}(t)$ (or $ Y(t)$)  from the  nonrelativistic potential
$Y({\bf r}')$ using the standard Scar\"odinger equations (12) and 
the well known inverse scattering theory \cite{Chad,New,Suz}.


Summarizing the above  formulation of the $NN$ scattering problem,  
we see that in order to apply the inverse scattering
methods to the present field-theoretical equation (1) one must 
take into account the structure of the $NN$ potential.
The complete $NN$ potential consists of the nonlocal (on-mass-shell
pions exchange) part (Fig.1) and of the off-mass shell meson exchange
part ${\cal Y}(t)$ (Fig.2).
The contribution of the potential shown in  Fig.1 is  constructed from
the $\pi N$ vertex functions with one of the nucleons being off-mass
shell.
These vertex functions can be obtained from the $\pi N$ phase shifts
by using the dispersion relations. 
Therefore in the present formulation
the inverse scattering problem  is
reduced to the determination of the  ${\cal Y}(t)$ potential in the
equation (2). This is
because the other part of the $NN$ potential in the
equation (1) can be constructed from the $\pi N$ scattering amplitudes.
One can build the potential ${\cal Y}$
from the corresponding $NN$ phase shifts,
by using the inverse scattering methods for the nonrelativistic
potential $v({\bf r})$
 of the Schr\"odinger equation (12)
{\footnotemark}.

\footnotetext{ Unlike the nonrelativistic case, in the considered
relativistic formulation
${\bf r}$ does not have the physical meaning of the coordinate.
This variable should be treated as the auxiliary variable which
is conjugate to the momentum ${\bf p}$.  
In the Ref.\cite{KMS} ${\bf x}=i\sqrt{1+p^2/m^2}\partial/\partial {\bf p}$
was treated as the generators of translation in the Euclidean 
${\bf p}$-space with  the ${\bf x}^2-{\bf L}^2/m^2$ Casimir operator of
the Lorentz group. In this field-theoretical approach  the
relativistic analogue of the Fourier transformations
(the Shapiro transformation \cite{Shap}
with the complete set of the functions $\xi({\bf p,x})$) was used.
The final equation
have the form $H_o(2E_{\bf p}-H_o)<{\bf r}|\phi_{\bf p}>
=-1/2mV(r,E_{\bf p})<{\bf r}|\phi_{\bf p}>$ with the free Hamiltonian
$H_o=2m\ ch(i/m \partial/\partial r)+2i/r\ sh(i/m \partial/\partial r)  
+\triangle_{\theta,\phi}/mr^2\ exp(i/m \partial/\partial r)$.
}

\vspace{3mm}

The potential ${\cal Y}(t)$ is generated by the equal-time
anticommutation relation (A6a) and it is generally the function of the 
$t$-variable.
For the simplest Lagrangian (A5) ${\cal Y}(t)$ consist
of the terms, describing the exchange by the    
off-mass shell mesons:  $\pi,\sigma,\rho,\omega,...$
 (A6a) (Fig.2A) and of the  contact terms (A6b) (Fig.2B).
 Therefore the determination of the potential ${\cal Y}$ in the framework 
of the inverse scattering methods can help us
to clarify  the form of the general meson-nucleon Lagrngians. 
The formulation considered above can be  extended easily to
the  $\pi N-\pi N$ and  $\pi N-\gamma N$ scattering reactions.

The author is indebted to H.V. Geramb and A.A. Suzko for discussions
and for the interest to this work.

\vspace{5mm}

{\centerline{\bf APPENDIX}}

\vspace{5mm}

The well known expression for the $NN$ scattering $S$-matrix with the local 
meson  and the local nucleon  field operators $\Phi(x)$ and $\Psi(x)$
has the form \cite{BD,IZ} 

$$S_{N' N'\Leftarrow N N}\equiv 
<out;{\bf p'}_{1}{s'}_{1}{\bf p'}_{2}{s'}_{2}|
{\bf p}_{1}{s}_{1}{\bf p}_{2}{s}_{N2};in>=
<in;{\bf p'}_{1}{s'}_{1}{\bf p'}_{2}{s'}_{2}|
{\bf p}_{1}{s}_{1}{\bf p}_{2}{s}_{N2};in>+$$
$$
+{\cal P}_{1'2'}{\cal P}_{12}
(2\pi)^4\ i\ \delta^{(4)}(p'_{1}+p'_{2}-p_{1}-p_{2})
{\cal A}_{N' N'\Leftarrow N N}\eqno(A1)$$

$${\cal A}_{N' N'\Leftarrow N N}={\cal P}_{1'2'}
<out;{\bf p'}_{1}{s'}_{1}|
J_{{\bf p'}_2{s'}_{2}}(0)|{\bf p}_{1}{s}_{1}{\bf p}_{2}{s}_{1};in>=
equal\ time\ anticommutators\ +
$$
$$
{\cal P}_{1'2'}{\cal P}_{12}\int d^4x e^{(-ip_{1}x)}
<{\bf p'}_1{s'}_{1}|\biggl(J_{{\bf p'}_2{s'}_{2}}(0)\theta(-x_o)
{\overline J}_{{\bf p}_1{s}_{1}}(x)\ -
\ {\overline J}_{{\bf p}_1{s}_{1}}(x)\theta(x_o)
J_{{\bf p'}_2{s'}_{2}}(0)\biggr)
|{\bf p}_{N2}{s}_{2}>\eqno(A2)$$

where ${\bf p'}_i{s'}_{i}$ and ${\bf p}_i{s}_{i}$ denote the three
momentum and the spin of the nucleon, $i=1,2$ correspondingly in the
final and in the initial states,  
${\cal P}_{12}=1/2(1-{\widehat P}_{12})$ stands for the antisymmetrization 
operator with the nucleon transposition operator  ${\widehat P}_{12}$,
$J_{{\bf p'}_2{s'}_{2}}(x)\equiv {\overline u}({\bf p'}_{2}s'_2)J(x)
{\overline u}({\bf p'}_{2}s'_2)
\Bigl(i\gamma^{\mu}\partial_{\mu}-m_{N}\Bigr)\Psi(x)$ is the nucleon
source operator which is determined by the Dirac equation of motion,
${u}({\bf p}_{N}s)$ stands for the Dirac bispinor function and 
$\theta(x_o)=1$  if $x_o>0$ and $\theta(x_o)=0$ if $x_o<0.$   
is the well known step function.

Substituting the completeness condition $\sum_n |n;in><in;
n|={\widehat {\bf 1}}$
into eq. (A2) between the source operators,we obtain after integration over $x$

$${\cal A}_{N' N'\Leftarrow N N}=equal\ time\ anticommutators\ +
(2\pi)^3{\cal P}_{1'2'}{\cal P}_{12}$$
$$
\biggl(\sum_{n=d,NN,\pi NN,...}
<{\bf p'}_{1}{s'}_1|J_{{\bf p'}_2{s'}_{2}}(0)|n;in>
{{\delta({\bf p}_{1}+{\bf p}_{2}-{\bf P}_n)}\over
{E_{{\bf p}_{1}}+E_{{\bf p}_{2}}-P_n^o+io}}
<in;n|{\overline J}_{{\bf p}_1{s}_{1}}(0)|{\bf p}_{2}{s}_2>$$
$$-\sum_{m=\pi,\pi\pi,...}
<{\bf p'}_{1}{s'}_1|{\overline J}_{{\bf p}_1{s}_{1}}(0)|m;in>
{{\delta({\bf p'}_{1}-{\bf p}_{1}-{\bf P}_ m)}\over
{E_{{\bf p'}_{1}}-E_{{\bf p}_{1}}-P_m^o}}
<in; m|J_{{\bf p'}_2{s'}_{2}}(0)|{\bf p}_{2}{s}_2>
\biggr),\eqno(A3)$$

where $p_N\equiv\Bigl(E_{{\bf p}_{N}},{\bf p}_{N}\Bigr)=
\Bigl(\sqrt{{{\bf p}_{N}}^2+m_N^2},{\bf p}_{N}\Bigr)$ denotes the 
four momentum of the on-mass shell nucleon.
 
Comparing  equation (A2) with  equation (A3), we see,
that the time-ordering procedure in eq. (A2) is replaced by the 
linear  propagator
which consists from the energies of the outside and inside particles.
Besides in the equation (A2)only the sum of the three-momentums of the 
all intermediate particles  is conserved.
However, the main property of the present field-theoretical formulation 
is that it is the only one, where the  one variable 
vertex functions like $<{\bf p'}_N|{\overline J}_{\beta}(0)|{\bf p}_{\pi}>$
or $<{\bf p'}_N|j_{\pi'}(0){\bf p"}_N>$ (in the  
equal time anticommutators) are required as input functions for the
construction of the $NN$ potential.

The equal time anticommutators in equations (A2) and (A3) have the form

$$equal\ time\ anticommutators\equiv{\cal Y}(t)=
{\cal P}_{1'2'}{\cal P}_{12}
<{\bf p'}_{1}{s'}_1|\Bigl\{J_{{\bf p'}_2{s'}_2}(0),
b^{\dagger}_{{\bf p}_1s_1}(0)\Bigr\}
|{\bf p}_{2}s_2>,\eqno(A4)$$

where the operator
$b^{\dagger}_{{\bf p}_1s_1}(y_o)=\int d^3 y e^{-i{p}_{1}y}
\gamma_o{\overline {\Psi}}(y)u({\bf p}_1s_1)$ 
tends to the nucleon creation (annihilation) operator in the asymptotic region
$\lim_{x_0\to\pm \infty}{b^+}_{{\bf p}s}(x_o) \Longrightarrow
{b^+}_{{\bf p}s}(out\ or\ in)$ and satisfy the anticommutation relations  
$\Bigl\{b_{{\bf p'}s'}(x_o),{b^+}_{{\bf p}s}(x_o)\Bigr\} =
{{E_{{\bf p}} }/{m_N}}\delta_{s's}\delta({\bf p'}-{\bf p})$.

The exact form of the equal-time anticommutators (A4) can be derived using
the usual form of the meson-nucleon Lagrngians

$${\cal L}_{int}=g_{\sigma}{\overline \Psi}\Psi\sigma+
{{f_{\pi}}\over{m_{\pi}}}{\overline \Psi}\gamma_5\gamma_{\mu}\Psi
\partial^{\mu}\Phi_{\pi}
+g_{V}{\overline \Psi}\gamma_{\mu}\Psi V^{\mu}
+{{f_{V}}\over{4m_N}}
{\overline \Psi}\sigma_{\mu\nu}\Psi(\partial^{\mu}V^{\nu}-
\partial^{\nu}V^{\mu}),\eqno(A5)$$
where  $\Phi_{\sigma}$, $\Phi_{\pi}$ and $\Phi_V$
denote the field operators of the  
$\sigma$, $\pi$  and of the $V=\rho,\omega$  mesons.
Using the equal-time commutation relation between the Heisenberg
field operators, we obtain

$${\cal Y}(t)\equiv{\cal P}_{1'2'}{\cal P}_{12}
<{\bf p'}_{1}{s'}_1|\Bigl\{J_{{\bf p'}_2{s'}_2}(0),
b^{\dagger}_{{\bf p}_1s_1}(0)\Bigr\}
|{\bf p}_{2}s_2>=
{\cal P}_{1'2'}{\cal P}_{12}\biggl($$
$$g_{\sigma}{\overline u}({\bf p'}_{2}{s'}_2) u({\bf p}_{1}s_1)
{{<{\bf p'}_{1}{s'}_1|j_{\sigma}(0)|{\bf p}_{2}s_2>}\over{t-m_{\sigma}^2}}  
+i{{f_{\pi}}\over{2m_Nm_{\pi}}}
{\overline u}({\bf p'}_{2}{s'}_2)\gamma_5 u({\bf p}_{1}s_1)
{{<{\bf p'}_{1}{s'}_1|j_{\pi}(0)|{\bf p}_{2}s_2>}\over{t-m_{\pi}^2}}$$
$$+
g_V{\overline u}({\bf p'}_{2}{s'}_2)\gamma_{\mu} u({\bf p}_{1}s_1)
{{<{\bf p'}_{1}{s'}_1|j_{V}^{\mu}(0)|{\bf p}_{2}s_2>}\over{t-m_{V}^2}}\biggr)
+contact\ terms\eqno(A6a)$$  
$$contact\ terms=
{\cal P}_{1'2'}{\cal P}_{12}\biggl(
{{E_{{\bf p'}_2}-E_{{\bf p}_1}}\over{2m_N}}
{\overline u}({\bf p'}_{2}{s'}_2)i{{f_{\pi}}\over{2m_Nm_{\pi}}}
\gamma_5\gamma_o u({\bf p}_{1}s_1)
{{<{\bf p'}_{1}{s'}_1|j_{\pi}(0)|{\bf p}_{2}s_2>}\over{t-m_{\pi}^2}}$$
$$+{{f_{V}}\over{8m_N}}
{\overline u}({\bf p'}_{2}{s'}_2)u({\bf p}_{1}s_1)
<{\bf p'}_{1}{s'}_1|{\overline \Psi}(0)\Psi(0)|{\bf p}_{N}s_2>$$
$$+{{f_{V}}\over{8m_N}}
{\overline u}({\bf p'}_{2}{s'}_2)\gamma_{5}\gamma_o u({\bf p}_{1}s_1)
<{\bf p'}_1{s'}_1|{\overline \Psi}(0)\gamma_{5}\gamma_o\Psi(0)|{\bf
p}_{2}s_2>
+...\biggr)\eqno(A6b)$$
where $t=(p'_N-p_N)^2\equiv (E_{{\bf p'}_N}-E_{{\bf p}_N})^2-
({\bf p'}_N-{\bf p}_N)^2$ and 
due to the Lorentz-covariance of the scalar ($\sigma$),
pseudoscalar ($\pi$) and vector ($V=\rho,\omega$) vertex functions
 the following simple expressions
$$<{\bf p'}_{N}|j_{\sigma}(0)|{\bf p}_{N}>=
g_{\sigma}G_{\sigma}(t){\overline u}({\bf p'}_{N}){u}({\bf p}_{N});\ \ \
<{\bf p'}_{N}|j_{\pi}(0)|{\bf p}_{N}>=
ig_{\pi}G_{\pi}(t){\overline u}({\bf p'}_{N})\gamma_{5}{u}({\bf
p}_{N})$$
$$
<{\bf p'}_{N}|j_{V}^{\mu}(0)|{\bf p}_{N}>=
g_{V}G_V(t){\overline u}({\bf p'}_{N})\gamma^{\mu}{u}({\bf p}_{N}).\eqno(A7)$$

are vaid.
The diagrammatic representations of the three-dimensional equations
(A6a) and (A6b) are given in the Fig. 2.
The first three terms of the eq. (A6a) corresponds to the one
off-mass shell meson exchange $NN$ interaction potential and these terms
exactly coincides with the OBE $NN$ Bonn potential.
However, in the original
derivation of the Bonn OBE potential model from quasipotential
equation, the dependence of meson-nucleon vertices on the off mass-shell
variables ${p_N}^2\ne m_N^2$ and ${{p'}_N}^2\ne m_N^2$ has been neglected.
In the present field-theoretical formulation this approximation is not needed.
In this meaning  the present derivation can be considered as an additional
justification for the Bonn OBE potential model of the $NN$ interaction.
The contact or overlapping terms in  eq. (A6b) are generated by
the nonrenormalizable pseudoscalar and vector parts in the
Lagrangian (A5). If we take the renormalizable pseudoscalar coupling
$L_{ps}=ig_{\pi}{\overline \Psi}\gamma_5\Psi$
instead  of the pseudovector coupling
in the Lagrangian (A5), the first term  in the right hand side of
eq. (A6b) vanishes \cite{MCH,M1}. The other terms of the relation (A6b) are  
produced by the nonrenormalizable fourth term of the Lagrangian
(A5). In  Ref. \cite{M1,MBFE} 
the structure of the contact (overlapping) terms of the 
$NN$ interaction potential  with quark degrees of freedom was investigated. 
 The contact terms of the $NN$ potential were shown to consist  of
quark-gluon exchange contributions. 
In addition  it was obtained 
that due to the structure of the equal-time commutators
the quark-gluon exchange terms do not violate the unitarity condition
for the $NN$ scattering amplitude.

The considered field-theoretical equations are connected analytically  
with  all the other field-theoretical equations, i. e. we can derive 
the Bethe-Salpeter equation 
in the framework of the $S$-matrix reduction technique.
Therefore, all results obtained
in the framework of this time-ordered three-dimensional  equations remain
valid in  the other field-theoretical approaches as well.
This formulation is free from the "three-dimensional ambiguities" 
which emerge during the 
reduction of the Bethe-Salpeter equation in the three dimensional form.
The  structure  of the present field-theoretical equations
does not depend
on the choice  of the form of the effective Lagrangian.

The second term of the eq. (A3) is depicted in the Fig.1a 
as the on-mass shell meson exchange diagram.
After the cluster decomposition procedure i.e. after separation
of connected and disconnected parts in the amplitudes
$<in; m|J_{{\bf p'}{s'}}(0)|{\bf p}{s}>$ and 
 $<in;n|{\overline J}_{{\bf p}s}(0)|{\bf p}s>$, one obtain the 
$8$ skeleton diagrams for the connected parts of the
transition amplitudes.
We can omit the diagrams with the two and more on-mass shell meson 
exchange and we will neglect the contributions of the $\pi d$ and $\pi NN$
intermediate states, and of the    
anti-particle ${\overline d},{\overline N}{\overline N},...$
intermediate states in the low energy region. Therefore,
after cluster decomposition  we  obtain
only other time time-ordered pi-meson exchange diagram which is
depicted in the Fig.1b. Thus we can define the inhomogeneous
term of the eq. (A3) as

$$
<{\bf p'}_1{s'}_1{\bf p'}_2{s'}_2|W|{\bf p}_1s_1{\bf p}_2s_2>=
equal\ time\ anticommutators
$$
$$
+(2\pi)^3{\cal P}_{1'2'}{\cal P}_{12}\biggl(
-\sum_{m=\pi,\pi\pi,...}
<{\bf p'}_{1}{s'}_1|{\overline J}_{{\bf p}_1{s}_{1}}(0)|m;in>_c
{{\delta({\bf p'}_{1}-{\bf p}_{1}-{\bf P}_ m)}\over
{E_{{\bf p'}_{1}}-E_{{\bf p}_{1}}-P_m^o}}
<in; m|J_{{\bf p'}_2{s'}_{2}}(0)|{\bf p}_{2}{s}_2>_c$$
$$+\sum_{m=\pi,\pi\pi,...}
<0|{\overline J}_{{\bf p}_1{s}_{1}}(0)|{\bf p}_{2}{s}_2\ m;in>
{{\delta(-{\bf p}_{2}-{\bf p}_{1}-{\bf P}_ m)}\over
{-E_{{\bf p}_{2}}-E_{{\bf p}_{1}}-P_m^o}}
<in; m\ {\bf p'}_{1}{s'}_1|J_{{\bf p'}_2{s'}_{2}}(0)|0>
\biggr),\eqno(A6c)$$
where the subscript $'c''$ denotes the connected part of the
corresponding transition amplitude.

Using the eq. (A6c)  we can rewrite the equation (A3) 
in the c.m. frame as

$$<{\bf p}'|T|{\bf p}>=<{\bf p}'|W|{\bf p}>
+\sum_d <{\bf p}'|T|{ P}_d>{1\over{E({\bf p})-m_d}}
<{P}_d|T^{\dagger}|{\bf p}>$$
$$+\int  <{\bf p}'|T|{\bf p}''>
{{d{\bf p}''}\over{E({\bf p})+io-E({\bf p}'')}}
<{\bf p}''|T^{\dagger}|{\bf p}>
\eqno(A8)$$
where  we  have omitted the spin variables for the sake of simplicity,
 $E({\bf p})=2\sqrt{{\bf p}^2+m_N^2}$ is the energy of the
$NN$ state,  $P_d=(m_d,{\bf 0})$ is the four momentum of deuteron in
the c.m. frame

$$<{\bf p}'|T|{\bf p}>\equiv
-<{\bf p'}_{1}{s'}_1|J_{{\bf p'}_2{s'}_{2}}(0)
|{\bf p}_{1}{s}_1{\bf p'}_{1}{s'}_1{\bf p}_{2}{s}_2;in>\eqno(A9a)$$  
$$<{\bf p}'|T|{ P}_d>\equiv
-<{\bf p'}_{1}{s'}_1|J_{{\bf p'}_2{s'}_{2}}(0)|{\bf P}_{d}{S}_d;in>
\eqno(A9b)$$

The potential term $V$ in  eq. (A8) consist of the  on-mass shell pion
exchange diagrams (see the second term of eq. (A3) and the diagrams
shown in Fig.1) and of the equal-time commutator (A4). 
In the general case the commutator
 is  reduced to the
off-mass shell 
$\pi$,$\sigma$,$\rho$,$\omega$-meson  exchange diagrams 
and to the contact terms shown in Fig.2A,B.
In the framework of the simplest Lagrangian (A5)
the exact form of this commutators 
is given by  eqs.(A6a,b).

$$W=V(on\ mass\ shell\ meson \ exchange)+
{\cal Y}(t)(off\ mass\ shell\ meson\ exchange)\eqno(A10)$$

The procedure of
the exact linearization of the quadratically nonlinear
 equations (A8) for  $\pi N$,
$NN$ and $\gamma N-\pi N-\gamma \pi N -\pi \pi N$  
scattering reactions is described in the ref.\cite{MCH,M1,MR,Mm3}.

The on-mass shell meson exchange part of the $NN$ potential $V$
(Fig.1) is nonhermitian, but ${\cal Y}^{\dagger}(t)={\cal Y}(t)$.
Nevertheless we can obtain 
the linear energy dependent potential
$<{\bf p}'|U(E)|{\bf p}>$ 
from  $<{\bf p}'|W|{\bf p}>$ 
$$<{\bf p}'|U\Bigl(E=E({\bf p}')\Bigr)|{\bf p}>=<{\bf p}'|W|{\bf p}>
\eqno(A11)$$
so that off energy shell $U(E)$ is hermitian
$<{\bf p}'|U^{\dagger}(E)|{\bf p}>=<{\bf p}'|U(E)|{\bf p}>$, 
but on the half-energy shell $U(E)$ (as well as $W$) is not hermitian
$<{\bf p}'|U^{\dagger}(E=E({\bf p}')|{\bf p}>\ne
<{\bf p}'|U(E)|{\bf p}>$.
This property allows us to derive the Lippmann-Schwinger type
equation  exactly , by using Eq. (A8) \cite{MCH,M1} exactly

$$
<{\bf p}'|t(E)|{\bf p}>=<{\bf p}'|U(E)|{\bf p}>
+\int<{\bf p}'|U(E)|{\bf p}''>
{{d{\bf p}''}\over{E+io-E({\bf p}'')}}
<{\bf p}''|t(E)|{\bf p}>\eqno(A12)$$

On the  energy shell the solution of equations (A8) and (A12)
coincide, i.e.
$<{\bf p}'|t\Bigl(E({\bf p})=E({\bf p}')\Bigr)|{\bf p}>=<{\bf p}'|T|{\bf p}>$.
Equation (A12) coincide with the equation (1) on the half-energy
shell $E=E({\bf p})$.

\medskip
\par

\end{document}